\documentclass[preprint,showpacs,preprintnumbers,amsmath,amssymb]{revtex4}

\def\undersim#1{\setbox9\hbox{${#1}$}{#1}\kern-\wd9\lower
    2.5pt \hbox{\lower\dp9\hbox to \wd9{\hss $_\sim$\hss}}}
\usepackage{amssymb}
\usepackage{amsmath}
\usepackage{graphicx}

\def\undersim#1{\setbox9\hbox{${#1}$}{#1}\kern-\wd9\lower
    2.5pt \hbox{\lower\dp9\hbox to \wd9{\hss $_\sim$\hss}}}

\def\mr{{\mathbf r}}

\def\mr{{\mathbf r}}

\def\mk{{\mathbf k}}

\begin{document}

\title{Squeezed back-to-back correlations of $K^+$$K^-$ in d+Au collisions at $\sqrt{s_{NN}}=200$ GeV and Au+Au collisions at $\sqrt{s_{NN}}=62.4$ GeV}

\author{Yong Zhang$^{1,\,2}$\footnote{zhy913@jsut.edu.cn}}
\author{Jing Yang$^{3,}$\footnote{yangjdut@gmail.com}}
\author{WeiHua Wu$^{1}$}
\affiliation{\small$^1$School of Mathematics and Physics,
Jiangsu University of Technology, Changzhou, Jiangsu 213001, China\\
$^2$School of Science,Inner Mongolia University of Science $\&$ Technology, Baotou,Inner Mongolia Autonomous Region 014010, China\\
$^3$School of Physics and School of International Education Teachers, Changchun Normal University, Changchun, Jilin 130032, China
   }

\date{\today}

\begin{abstract}
We investigate the squeezed back-to-back correlations (BBC) of $K^+$$K^-$, caused by the mass
modification of the particle in the dense medium formed in d+Au collisions at
$\sqrt{s_{NN}}=200$ GeV and Au+Au collisions at $\sqrt{s_{NN}}=62.4$ GeV.
Considering some kaons may be not affected by the medium,
we further study the BBC functions of $K^+$$K^-$ when parts of all kaons have a mass-shift. Our results indicate that
the BBC functions of $K^+$$K^-$ can be observed when only about 10$\%$ of all kaons have a mass-shift
in d+Au collisions at $\sqrt{s_{NN}}=200$ GeV and the peripheral collisions of
Au+Au at $\sqrt{s_{NN}}=62.4$ GeV. Since the BBC function is caused by the mass-shift due to the
interaction between the particle and the medium, the successful detection of the BBC function
indirectly marks that the dense medium has formed in these collision systems. We suggest to
measure the BBC function of $K^+$$K^-$ experimentally in d+Au collisions at $\sqrt{s_{NN}}=200$ GeV
and peripheral collisions of Au+Au at $\sqrt{s_{NN}}=62.4$ GeV.

%Keywords: squeezed back-to-back correlations,$K^+$$K^-$, mass modification, dense medium.

\end{abstract}

\maketitle

\section{Introduction}

In high-energy heavy-ion collisions, people study the whole collision processes and the
properties of matter produced in the early stage of collisions by analyzing the observables
of particles in final state. The interactions within the dense medium before kinetic freeze-out
may affect the experimental observables and dilute the early signals, so the study of interactions
between the particle and the dense medium has always been a topic of
concern \cite{Ko1992,Asakawa1994,Fuchs04,Fuchs06,He2011,He2014,Cao2015}.
In the late 1990s, M. Asakawa $et$ $al$. proposed that particle mass might be modified by
the interaction between the particle and the medium formed in high-energy heavy-ion
collisions, and therefore lead to a squeezed back-to-back correlation (BBC) of boson-antiboson \cite{AsaCso96,AsaCsoGyu99}. This BBC is the consequences of a quantum mechanical transformation
relating in-medium quasi-particles to the two-mode squeezed states of their free observable
counterparts, through a Bogoliubov transformation between the creation (annihilation) operators
of the quasi-particles and the free observable particles \cite{AsaCso96,AsaCsoGyu99,Padula06}.
Since the BBC is related to the dense source, the investigations of the BBC may provide
a new sight for people to understand the interactions between the particle and the dense medium
formed in high-energy heavy-ion collisions.

The BBC function is defined as \cite{AsaCso96,AsaCsoGyu99}
\begin{equation}
\label{BBCf}
C(\mk,-\mk) = 1 + \frac{|G_s(\mk,-\mk)|^2}{G_c(\mk,\mk) G_c(-\mk,-\mk)},
\end{equation}
where $G_c(\mk_1,\mk_2)$ and $G_s(\mk_1,\mk_2)$ are the chaotic and squeezed
amplitudes, respectively. For hydrodynamic sources, with the formula derived by
Makhlin and Sinyukov \cite{MakhSiny,Siny}, the chaotic and squeezed amplitudes
can be expressed as \cite{AsaCsoGyu99,Padula06,Padula10,YZHANG15a,YZHANG16}
\begin{eqnarray}
\label{Gchydro}
&& G_c({\mk_1},{\mk_2})\!=\!\int \frac{d^4\sigma_{\mu}(r)}{(2\pi)^3}
K^\mu_{1,2}\, e^{i\,q_{1,2}\cdot r}\,\! \Bigl\{|c'_{\mk'_1,\mk'_2}|^2\,
n'_{\mk'_1,\mk'_2}~~~~~~\nonumber \\
&& \hspace*{19mm}
+\,|s'_{-\mk'_1,-\mk'_2}|^2\,[\,n'_{-\mk'_1,-\mk'_2}+1]\Bigr\},
\end{eqnarray}
\begin{eqnarray}
\label{Gshydro}
&& G_s({\mk_1},{\mk_2})\!=\!\int \frac{d^4\sigma_{\mu}(r)}{(2\pi)^3}
K^\mu_{1,2}\, e^{2 i\,K_{1,2}\cdot r}\!\Bigl\{s'^*_{-\mk'_1,\mk'_2}
c'_{\mk'_2,-\mk'_1}~~~~~\nonumber \\
&& \hspace*{18mm}
\times n'_{-\mk'_1,\mk'_2}+c'_{\mk'_1,-\mk'_2} s'^*_{-\mk'_2,\mk'_1}
[n'_{\mk'_1,-\mk'_2} + 1] \Bigr\}.
\end{eqnarray}
Where $d^4\sigma_{\mu}(r)$ is the four-dimension element of freeze-out
hypersurface, which can be determined by hydrodynamic source for a fixed
freeze-out temperature. $q^{\mu}_{1,2}=k^{\mu}_1-k^{\mu}_2$, $K^{\mu}_{1,2}=
(k^{\mu}_1+k^{\mu}_2)/2$, and $\mk_i'$ is the local-frame momentum
corresponding to $\mk_i~(i=1,2)$.  The quantities $c'_{\mk'_1,\mk'_2}$ and $s'_{\mk'_1,
\mk'_2}$ are the coefficients of Bogoliubov transformation between
the creation (annihilation) operators of the quasiparticles and the
free particles, and $n'_{\mk'_1,\mk'_2}$ is the boson distribution
associated with the particle pair \cite{AsaCso96,AsaCsoGyu99,Padula06,Padula10,YZHANG15a,YZHANG16}.
$s'_{\mk'_1,\mk'_2}$ will be zero if there is no mass modification, and the BBC function
 will be 1. In Eq. (\ref{Gshydro}), the factor $e^{2iK_{1,2}\cdot r}$ is equal to
$e^{2i\omega_{\mk}t}$ for $\mk_1=\mk$, $\mk_2=-\mk_1=-\mk$. So the BBC
function $C(\mk,-\mk)$ is sensitive to the temporal distribution of
the freeze-out points \cite{AsaCsoGyu99,Padula06,Padula10,YZHANG15a,YZHANG16,Kno11}, which
will be very large for a narrow temporal distribution of the particles' freeze-out points \cite{AsaCsoGyu99,Padula06,Padula10,YZHANG16}.

In experiment, the first search of BBC signals was performed by the PHENIX collaborations \cite{NAGY},
the $K^+$$K^-$ BBC function is only 1$\%$ larger than one in the central Au+Au collisions at
$\sqrt{s_{NN}}=200$ GeV, not present on a significant level.
The simulation results indicate that the $K^+$$K^-$ BBC function is suppressed by
the wide temporal distribution of the particles' freeze-out points and cannot be
detected in Au+Au collisions at $\sqrt{s_{NN}}=200$ GeV for central collisions \cite{YZHANG16}.

Recently, the long-range angular correlations has been first detected  by the
CMS Collaboration in $pp$ collisions at 7 TeV at the LHC\cite{CMS-JHEP2010}, and later it was also observed in $pp$ collisions \cite{ATLAS-PRL2016,CMS-PRL2016} and p+Pb collisions
\cite{CMS-PLB2013,ALICE-PLB2013,ATLAS-PRL2013,LHCb-PLB2016} by other collaborations at the LHC.
This effect and the elliptic anisotropy of inclusive and identified
hadrons were also successfully detected by the PHENIX collaboration in the 5$\%$ most central d+Au
collisions at $\sqrt{s_{NN}}=200$ GeV at the RHIC \cite{PHENIX-PRL13,PHENIX-PRL15}.
It indicates that the hydrodynamically expanding quark-gluon plasma (QGP) medium may
be created in the small systems, and hydrodynamic calculations are in good agreement
with measured $v_2$ in small systems \cite{Bzdak-PRC13,GuangYou-PRC14,Romatschke15,Bozek15,PHNEIX-PRC17}.
The BBC function is sensitive to the temporal distribution of the particles' freeze-out points,
and the narrow temporal distribution corresponding to a small hydrodynamic source may lead to
a large BBC. Motivated by this, we study the BBC function of $K^+$$K^-$ in d+Au collisions
at $\sqrt{s_{NN}}=200$ GeV by using the VISH2$+$1 code \cite{VISH2+1-1,VISH2+1-2} to
simulate the evolutions of the particle-emitting sources. And we also calculate the
BBC functions in Au+Au collisions at $\sqrt{s_{NN}}=62.4$ GeV.
The event-by-event initial conditions of MC-Glb \cite{VISHb} are employed at $\tau_0=0.6$ fm/$c$
in the simulations, and the ratio of the shear viscosity to entropy density of the QGP
is taken to be 0.08 \cite{Shen11-prc,Qian16-prc}.

The rest of this paper is organized as follows. In Sec. II, we present the transverse momentum
spectra of kaon, and the spatio-temporal properties of kaon emission source. In Sec. III,
we calculate the BBC functions of $K^+$$K^-$ for central and peripheral collisions of d+Au at
$\sqrt{s_{NN}}=200$ GeV and Au+Au at $\sqrt{s_{NN}}=62.4$ GeV. Considering some particles may be not affected
by the medium in the collision, we further introduce a gaussian factor to describe the
probability of the particles which affected by the medium and study the BBC functions of $K^+$$K^-$
when parts of all kaons are affected by the medium in this section. Finally, a summary and conclusions of this paper are given in Sec. IV.

\section{Spectra and emission source of Kaon}
\begin{figure}[htbp]
\includegraphics[scale=0.68]{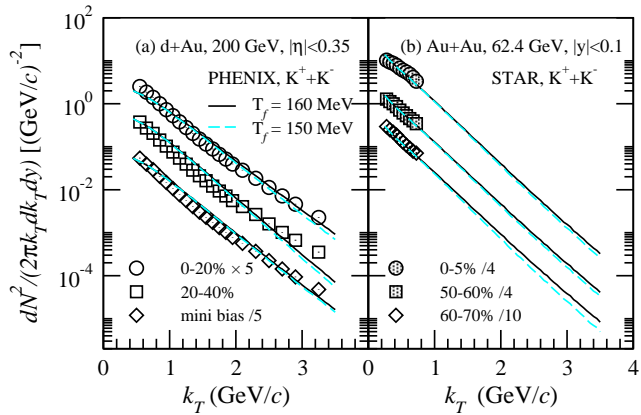}
\vspace*{-2mm}
\caption{Transverse momentum spectra of kaon calculated with VISH2+1 for d+Au collisions at
$\sqrt{s_{NN}}=200$ GeV (left panel) and Au+Au collisions at $\sqrt{s_{NN}}=62.4$ GeV (right panel). The experimental
data of d+Au collisions are measured by the PHENIX Collaboration \cite{PHENIX_spectra},
and the experimental data of Au+Au collisions are from the STAR Collaboration
measurements \cite{STAR-PRC09}. The solid lines are the simulated spectra for freeze-out
temperature $T_f = 160$ MeV, and the dashed lines are the simulated spectra for $T_f = 150$ MeV.}
\label{fig1}
\end{figure}

In Fig. \ref{fig1}, we show the
transverse momentum spectra of kaon simulated by the viscous hydrodynamic code for the collisions
in different centrality ranges, and the experimental data \cite{PHENIX_spectra,STAR-PRC09}
are also plotted. One can see that the simulated spectra for freeze-out temperature $T_f = 160$ MeV suit the experimental data better than for $T_f = 150$ MeV.
So the freeze-out temperature $T_f$ of kaon will be selected as 160 MeV in this work.

For hydrodynamic sources with a Bjorken cylinder, the four-dimensional element of the
freeze-out hypersurface can be written as
\begin{equation}
d^4\sigma_{\mu}(r)=f_{\mu}(\tau, \mr_{\perp}, \eta)\, d\tau d^2
\mr_{\perp} d\eta,
\end{equation}
where $\tau$, $\mr_{\perp}$, and $\eta$ are the proper time, transverse
coordinate, and space-time rapidity of the element.  The function $f_{\mu}
(\tau,\mr_{\perp},\eta)$ is related to the freeze-out mechanism that is
considered, and $K^{\mu}_{1,2}f_{\mu}(\tau,\mr_{\perp},\eta)$ corresponds
to the source distributions of proper time and space in the calculations
[see Eqs. (\ref{Gchydro}) and (\ref{Gshydro})].
\begin{figure}[htbp]
\vspace*{5mm}
\includegraphics[scale=0.5]{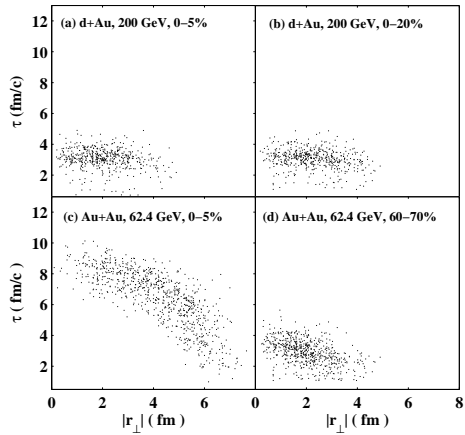}
\vspace*{-3mm}
\caption{Distributions of the freeze-out points of kaon in the $z=0$ plane for d+Au
collisions at $\sqrt{s_{NN}}=200$ GeV (top panel) and Au+Au collisions at $\sqrt{s_{NN}}=62.4$ GeV (bottom panel).}
\label{fig2}
\end{figure}

\begin{figure}[htbp]
\vspace{5mm}
\includegraphics[scale=0.5]{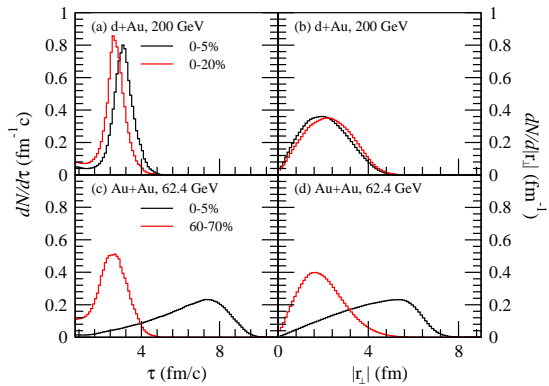}
\vspace{-2mm}
\caption{Normalized distributions of time and transverse coordinate of
kaon freeze-out points in the $z=0$ plane for d+Au collisions at $\sqrt{s_{NN}}=200$ GeV (top panel)
and Au+Au collisions at $\sqrt{s_{NN}}=62.4$ GeV (bottom panel) as in Fig. \ref{fig2}.}
\label{fig3}
\end{figure}
In Fig. \ref{fig2} (a) and (b), we show the space-time distributions of the freeze-out
points, $k^{\mu}f_{\mu}(\tau,\mr_{\perp},\eta)$, of kaon in the $z=0$ plane in d+Au
collisions at $\sqrt{s_{NN}}=200$ GeV for two centrality ranges. The spatial distribution
is wide, and the temporal distribution is narrow for d+Au collisions. The freeze-out points of kaon in
central and peripheral collisions of Au+Au at $\sqrt{s_{NN}}=62.4$ GeV are
plotted in Fig. \ref{fig2} (c) and (d), the width of the distributions increases with
decreasing collision centrality. In Fig. \ref{fig3}, we also show the normalized distributions of time
and transverse coordinate of kaon freeze-out points in the $z =$ 0 plane of the collisions as in Fig. \ref{fig2}.

\begin{table}[htb]
\caption{Spatio-temporal properties of kaon emission source.}
\begin{tabular}{cccccc}
\hline\hline
~Collision systems&~Centrality&~~$\langle\,\bar{\tau}\,\rangle$\,(fm/\emph{c})&~~$\langle\,\overline{|r_{\bot}|}\,\rangle$\,(fm)~~&
~~$\langle\,{\sigma_{\tau}}\,\rangle$\,(fm/\emph{c})~~&$\langle\,{\sigma_{r}}\,\rangle$\,(fm)\\
\hline
d+Au (\,200 GeV\,)~&~0-5$\%$~&~3.02~&~2.14&~0.66&~0.91\\
d+Au (\,200 GeV\,)~&~0-20$\%$~&~2.59~&~2.31&~0.58&~0.87\\
d+Au (\,200 GeV\,)~&~20-40$\%$~&~2.03~&~2.62&~0.47&~0.75\\
Au+Au (\,62.4 GeV\,)~&~0-5$\%$~&~6.27~&~4.19&~1.87&~1.58\\
Au+Au (\,62.4 GeV\,)~&~50-60$\%$~&~2.87~&~2.20&~0.84&~0.99\\
Au+Au (\,62.4 GeV\,)~&~60-70$\%$~&~2.46~&~2.02&~0.71&~0.93\\
\hline\hline
\end{tabular}
\label{Tab-1}
\end{table}

In table \ref{Tab-1}, we show some spatio-temporal properties of kaon emission source.
Where $\bar{{\tau}}$ and $\overline{|r_{\bot}|}$ are the average time and the average transverse radius of
kaon freeze-out points in the $z=0$ plane for a single event. $\sigma_{\tau}$ and $\sigma_{r}$
are the standard deviation of the freeze-out time and transverse radius for a single event:
\begin{eqnarray}
\label{Sigmat}
\sigma_{\tau} = \sqrt{\frac{1}{N}\sum_{i=1}^{N}(\,{\tau}_{i}-\bar{{\tau}}\,)^{\,2}}
\end{eqnarray}
\begin{eqnarray}
\label{Sigmat}
\sigma_{r} = \sqrt{\frac{1}{N}\sum_{i=1}^{N}(\,{|{r_\bot}|_i}-\overline{|r_{\bot}|}\,)^{\,2}}
\end{eqnarray}

Where $N$ is the total number of the freeze-out points for a single event, ${\tau}_{i}$ and ${|{r_\bot}|_i}$
are the space-time coordinate of the freeze-out point denoted by $i$. $\langle \cdots \rangle$ in
table \ref{Tab-1} means events average. The $\langle\,{\sigma_{\tau}}\,\rangle$ and $\langle\,{\sigma_{r}}\,\rangle$
are the average width of time and space of freeze-out points for many events, respectively.
They decrease with increasing collision centrality in d+Au
collisions at $\sqrt{s_{NN}}=200$ GeV and Au+Au collisions at $\sqrt{s_{NN}}=62.4$ GeV.

\section{BBC results}
The BBC function is caused by the mass-shift due to the interaction between the particle and the medium,
and it will be 1 if there is no mass-shift. In this work, the mass of kaon in medium is treated as a parameter, and
we assume that there is no mass-shift if the particle is not affected by the medium. The size of the systems of d+Au collisions and the peripheral Au+Au collisions are too small, and some particles may be not affected
by the medium in these collision events. Therefor, the BBC results of $K^+$$K^-$ will be shown in two subsections. In section A, we will show the BBC results for all kaons with mass-shift. In section B, the BBC results
for parts of all kaons with mass-shift will be shown.

\subsection{All kaons have a mass-shift}

When all particles have a mass-shift, the BBC function averaged
over event-by-event calculations for many events was defined as \cite{YZHANG16}
\begin{equation}
C(\mk,-\mk)=1+\frac{\sum_{i=1}^{N_E}|G_{si}(\mk,-\mk)|^2}{
\sum_{i=1}^{N_E}|G_{ci}(\mk,\mk)|^2},
\end{equation}
where $N_E$ is the total event number.

\begin{figure}[htbp]
\vspace*{5mm}
\includegraphics[scale=0.78]{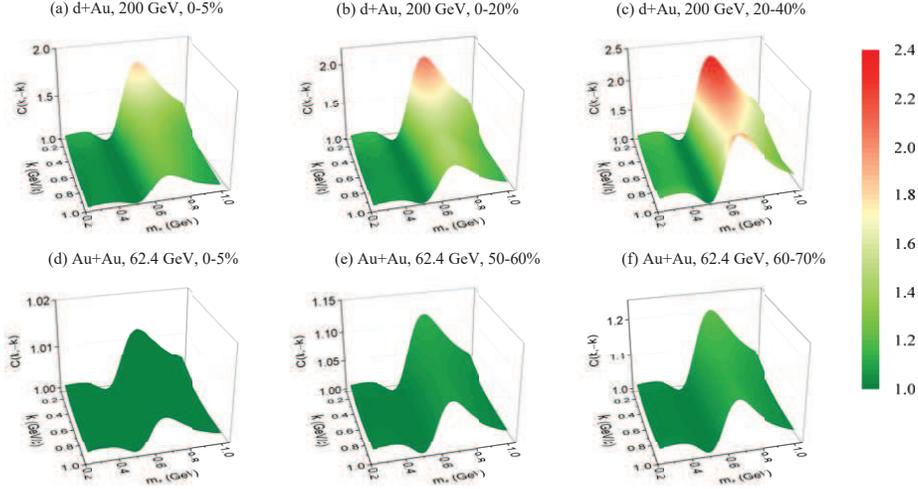}
\vspace*{-2mm}
\caption{BBC functions of $K^+$$K^-$ averaged over event-by-event calculations for central
and peripheral collisions of d+Au at $\sqrt{s_{NN}}=200$ GeV (top panel) and Au+Au at $\sqrt{s_{NN}}=62.4$ GeV (bottom panel).}
\label{fig4}
\end{figure}

In Fig. \ref{fig4}, we show the BBC functions of $K^+$$K^-$ averaged over event-by-event
calculations for central and peripheral collisions of d+Au at $\sqrt{s_{NN}}=200$ GeV and
Au+Au at $\sqrt{s_{NN}}=62.4$ GeV. The BBC function is shown in $k$$-$$m_*$ plane,
where $m_*$ is the kaon mass in the dense medium. If $m_*$ is equal to the kaon
mass in vacuum (494 MeV), there is no mass-shift, and the BBC functions of
$K^+$$K^-$ become one (see Fig. \ref{fig4}). For d+Au collisions, the
freeze-out temporal distribution of kaon is much narrow (see Fig. \ref{fig2} and Fig. \ref{fig3}), and the
temporal width parameter of the kaon emission source $\langle\,{\sigma_{\tau}}\,\rangle$ is also
very small (see table \ref{Tab-1}). As well known, a narrow temporal distribution may lead to
a large BBC, the BBC functions of $K^+$$K^-$ for the collisions of d+Au at $\sqrt{s_{NN}}=200$ GeV
are much stronger than that for the collisions of Au+Au at $\sqrt{s_{NN}}=200$ GeV and Pb+Pb at
$\sqrt{s_{NN}}=2.76$ TeV \cite{YZHANG16}, and it is very likely to be observed experimentally.
An observable signal of $K^+$$K^-$ BBC is also provided in peripheral collisions of Au+Au at
$\sqrt{s_{NN}}=62.4$ GeV. For hydrodynamic source, the BBC functions of $K^+$$K^-$
decrease with increasing the momentum $k$ for a fixed $m_*$. This phenomenon is different from the results
for the expanding source with a exponential decay emission time distribution
\cite{AsaCsoGyu99,Padula06,Padula10,YZHANG_CPC15,IJMPE}. This can be attributed in two reasons:
first, the BBC functions are affected by the emission time distribution of the particle
\cite{Padula10,YZHANG15a,YZHANG16}, and second, the anisotropic flow also affects the BBC functions \cite{IJMPE}
(e.g., in Ref.\cite{Padula10}, the BBC functions of $K^+$$K^-$ decrease with increasing the momentum $k$ for a fixed $m_*$ for a expanding source with the $\alpha$-stable L\'{e}vy emission time distribution).

\begin{figure}[htbp]
\vspace*{5mm}
\includegraphics[scale=0.66]{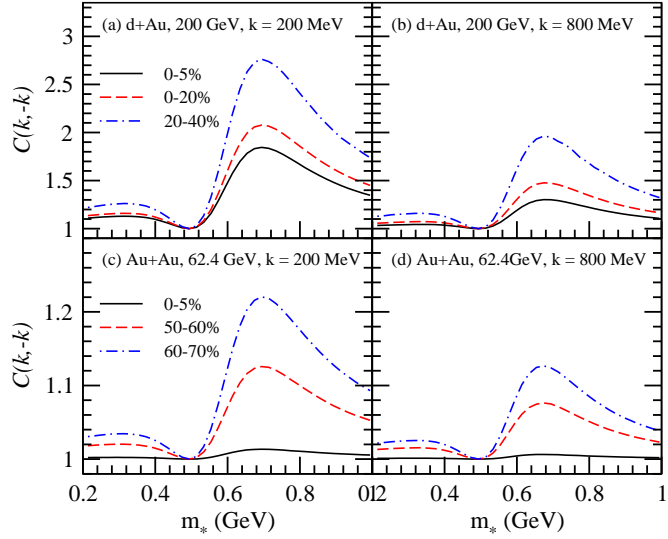}
\vspace*{-2mm}
\caption{BBC functions of $K^+$$K^-$ as a function of $m_*$ at $k = 200$ MeV and $k = 800$ MeV for
central and peripheral collisions of d+Au at $\sqrt{s_{NN}}=200$ GeV and Au+Au at $\sqrt{s_{NN}}=62.4$ GeV.}
\label{fig5}
\end{figure}
In Fig. \ref{fig5}, the BBC functions of $K^+$$K^-$ are shown as a function of
$m_*$ at $k = 200$ MeV and $k = 800$ MeV for central and peripheral collisions of d+Au at
$\sqrt{s_{NN}}=200$ GeV and Au+Au at $\sqrt{s_{NN}}=62.4$ GeV. For the same momentum, the BBC
function increases with the temporal width parameter $\langle\,{\sigma_t}\,\rangle$ decreases.

\subsection{Parts of all kaons have a mass-shift}

\begin{figure}[htbp]
\vspace*{5mm}
\includegraphics[scale=0.66]{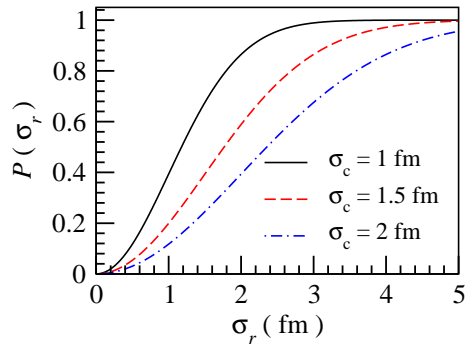}
\vspace*{-2mm}
\caption{Probability of the particles with mass-shift as a function of $\sigma_{r}$
for different spatial width cut parameter $\sigma_{c}$.}
\label{fig6}
\end{figure}

The narrow temporal distribution corresponding to a small source may lead to a large BBC.
But if the source is too small, the particles may be not affected by the medium. So we
introduce a gaussian factor $e^{-\sigma_{r}^2/2\sigma_{c}^2}$ to
describe the probability of the particles without mass-shift.
Where $\sigma_{c}$ is a spatial width cut parameter,
and the probability of the particles with mass-shift is $P(\sigma_r)\,=\,1-e^{-\sigma_{r}^2/2\sigma_{c}^2}$.
In Fig. \ref{fig6}, we show the probability $P(\sigma_r)$ as a function of $\sigma_{r}$ with different $\sigma_{c}$. For the source with a certain $\sigma_{r}$, the probability decreases with increasing the cut parameter $\sigma_{c}$.

The probability $P(\sigma_r)$ is fluctuating among the events for a fixed cut parameter $\sigma_{c}$, because
the width parameter $\sigma_{r}$ of the source is fluctuating among the events.
The average probability, $\langle\,P(\sigma_r)\,\rangle$, of the kaons with mass-shift for central and peripheral collisions of d+Au at $\sqrt{s_{NN}}=200$ GeV and Au+Au at $\sqrt{s_{NN}}=62.4$ GeV was shown in table \ref{Tab-2}, where $\langle \cdots \rangle$ means events average.

\begin{table}[htb]
\caption{The average probability of the kaons with mass-shift for central
and peripheral collisions of d+Au at $\sqrt{s_{NN}}=200$ GeV and Au+Au at $\sqrt{s_{NN}}=62.4$ GeV.}
\begin{tabular}{cccccc}
\hline\hline
~$\langle\,P(\sigma_r)\,\rangle$:&~$\sigma_{c}\,=\,1\,$fm&~~$\sigma_{c}\,=\,1.5\,$fm~~&
~~$\sigma_{c}\,=\,2\,$fm\\
\hline
d+Au\,(\,200 GeV\,)\,0-5$\%$~&~33.8\%~&~16.9\%&~9.9\%\\
d+Au\,(\,200 GeV\,)\,0-20$\%$~&~31.6\%~&~15.7\%&~9.2\%\\
\,d+Au\,(\,200 GeV\,)\,20-40$\%$~&~24.5\%~&~11.9\%&~6.9\%\\
Au+Au\,(\,62.4 GeV\,)\,0-5$\%$~&~71.3\%~&~42.6\%&~26.9\%\\
~\,Au+Au\,(\,62.4 GeV\,)\,50-60$\%$~&~38.9\%~&~19.9\%&~11.8\%\\
~\,Au+Au\,(\,62.4 GeV\,)\,60-70$\%$~&~34.9\%~&~17.6\%&~10.4\%\\
\hline\hline
\end{tabular}
\label{Tab-2}
\end{table}

When parts of all particles have a mass-shift, the BBC function
averaged over event-by-event calculations for many events becomes
\begin{eqnarray}
C(\mk,-\mk)=1+\frac{\sum_{i=1}^{N_E}(1-e^{-\sigma_{r}^2/2\sigma_{c}^2})|G_{si}(\mk,-\mk)|^2}
{\sum_{i=1}^{N_E}[(1-e^{-\sigma_{r}^2/2\sigma_{c}^2})|G_{ci}(\mk,\mk)|^2
+e^{-\sigma_{r}^2/2\sigma_{c}^2}|G_{ci}^{0}(\mk,\mk)|^2]},
\end{eqnarray}
where $G_{ci}$ and $G_{si}$ are the chaotic and squeezed amplitudes for one event which denoted
by $i$, respectively. And $G_{ci}^{0}$ is the chaotic amplitude when there is no mass shift.

\begin{figure}[htbp]
\vspace*{5mm}
\includegraphics[scale=0.66]{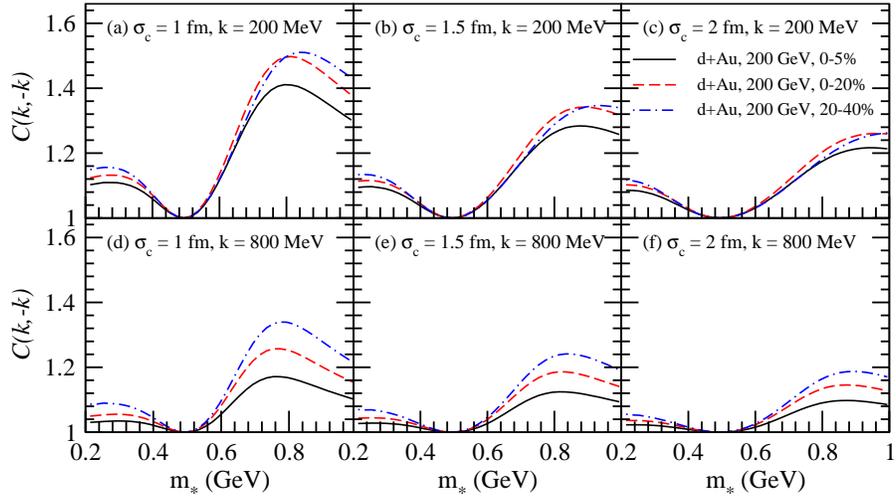}
\vspace*{-2mm}
\caption{BBC functions of $K^+$$K^-$ as a function of $m_*$ at $k = 200$ MeV (top panel) and $k = 800$ MeV (bottom panel)
with various cut parameters $\sigma_{c}$ in central and peripheral collisions of d+Au at
$\sqrt{s_{NN}}=200$ GeV.}
\label{fig7}
\end{figure}

\begin{figure}[htbp]
\vspace*{5mm}
\includegraphics[scale=0.66]{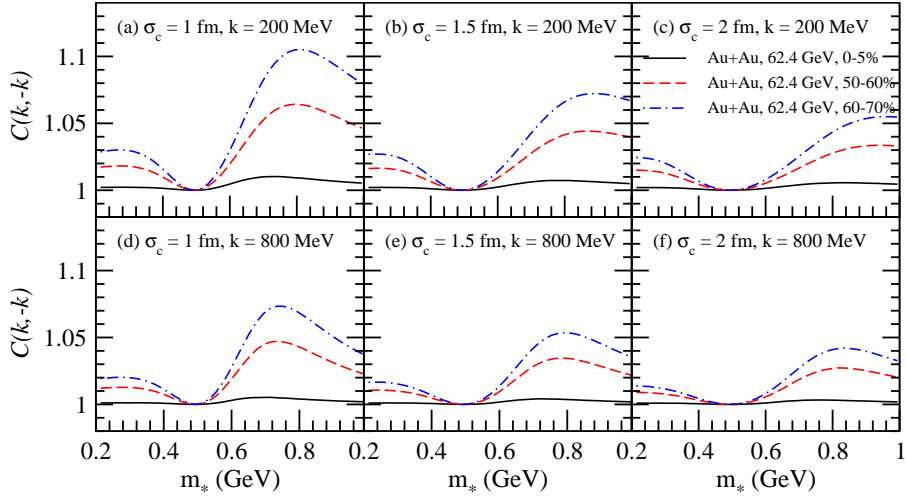}
\vspace*{-2mm}
\caption{BBC functions of $K^+$$K^-$ as a function of $m_*$ at $k = 200$ MeV (top panel) and $k = 800$ MeV (bottom panel)
with various cut parameters $\sigma_{c}$ in central and peripheral collisions of Au+Au at
$\sqrt{s_{NN}}=62.4$ GeV.}
\label{fig8}
\end{figure}

In Fig. \ref{fig7}, BBC functions of $K^+$$K^-$ are shown  as a function of $m_*$ at
$k = 200$ MeV and $k = 800$ MeV with various cut parameters $\sigma_{c}$ in central and
peripheral collisions of d+Au at $\sqrt{s_{NN}}=200$ GeV. For a certain cut parameter $\sigma_{c}$,
the probability of the kaons with mass-shift decreases with increasing the collision centralities (see table \ref{Tab-2}), so the BBC functions are more suppressed in peripheral collisions than in central collisions. A large
cut parameter $\sigma_{c}$ corresponds to a small probability of the kaons with mass-shift will
lead a small BBC (see table \ref{Tab-2} and Fig.\,\ref{fig7}). The BBC functions of $K^+$$K^-$ may be
observable when only 9.9$\%$ of all kaons have a mass-shift in 0-5$\%$ d+Au collisions at $\sqrt{s_{NN}}=200$ GeV for $\sigma_{c} = 2$ fm (see Fig. \ref{fig7} (c) and (f)).

In Fig. \ref{fig8}, we show the BBC functions of $K^+$$K^-$ as a function of $m_*$ at
$k = 200$ MeV and $k = 800$ MeV with various cut parameters $\sigma_{c}$ in central and
peripheral collisions of Au+Au at $\sqrt{s_{NN}}=62.4$ GeV. The BBC functions of $K^+$$K^-$
decrease with increasing the cut parameter $\sigma_{c}$, because the probability of the kaons with mass-shift
decreases with increasing the cut parameter. In peripheral
collisions of Au+Au at $\sqrt{s_{NN}}=62.4$ GeV, the BBC functions of $K^+$$K^-$ may perhaps
provide an observable signal.

\section{Summary and conclusions}
In high-energy heavy-ion collisions, the interactions between the particle and the dense medium may
lead to a modification of the particle mass, and thus give rise to a squeezed BBC of boson-antiboson.
The investigations of the BBC in previous works indicate that the BBC function $C(\mk,-\mk)$
is sensitive to the temporal distribution of the freeze-out points \cite{AsaCsoGyu99,Padula06,Padula10,YZHANG15a,YZHANG16,Kno11}. The BBC function will be very large for a narrow
temporal distribution of the particles' freeze-out points \cite{AsaCsoGyu99,Padula06,Padula10,YZHANG16}.

In this paper, we focus on studying the observability
of the BBC functions of $K^+$$K^-$ in d+Au collisions at $\sqrt{s_{NN}}=200$ GeV and Au+Au collisions at $\sqrt{s_{NN}}=62.4$ GeV by using the VISH2$+$1 code \cite{VISH2+1-1,VISH2+1-2} to simulate the evolutions of the particle-emitting sources. Compared with previous work \cite{YZHANG16}, we select the freeze-out temperature of kaon by comparing the simulated transverse momentum spectra of kaon with experimental data.
For all kaons have a mass-shift, the
BBC functions of $K^+$$K^-$ provide an observable signal in d+Au collisions at $\sqrt{s_{NN}}=200$ GeV and peripheral collisions of Au+Au at $\sqrt{s_{NN}}=62.4$ GeV. Considering that the systems of d+Au collisions and peripheral collisions of Au+Au are too small, the kaons may be not affected by the medium. Therefore, we introduce a gaussian factor to describe the probability of the kaons with mass-shift and further study the BBC functions of $K^+$$K^-$ when parts of all kaons with mass-shift. The BBC functions of $K^+$$K^-$ may be observable when only about 10$\%$ of all kaons have a mass-shift in d+Au collisions at $\sqrt{s_{NN}}=200$ GeV and peripheral collisions of Au+Au at $\sqrt{s_{NN}}=62.4$ GeV. Since the BBC function corresponds to the interactions between the particle and the dense medium, the successful detection of the BBC function may indirectly mark that the dense medium has formed in these collision systems. We suggest to measure the BBC function of $K^+$$K^-$ experimentally in d+Au collisions at $\sqrt{s_{NN}}=200$ GeV and peripheral collisions of Au+Au at $\sqrt{s_{NN}}=62.4$ GeV.

\begin{acknowledgments}
This research was supported by the National Natural Science Foundation
of China under Grant No. 11647166,11747155, the Natural Science Foundation of Inner Mongolia
under Grant No. 2017BS0104, Changzhou Science and Technology Bureau CJ20180054, and the Foundation of Jiangsu University of Technology
under Grant No. KYY17028.
\end{acknowledgments}

\end{document}